\documentclass[aps,prd,12pt,preprint,nofootinbib,tightenlines,floatfix,superscriptaddress,showpacs]{revtex4}
 
\usepackage{graphicx}% Include figure files
\usepackage{rotating}
\usepackage{dcolumn}% Align table columns on decimal point
\usepackage{bm}% bold math
\usepackage{longtable}% Include figure files

\begin{document}

\def\figwid{6.5in}

\newcommand{\ie}{{\it i.e.}}
\newcommand{\etal}{{\it et al.}}
\newcommand{\eg}{{\it e.g.}}
\newcommand{\etap}{{\eta\,'}}
\newcommand{\psip}{{\psi(2S)}}
\newcommand{\dilep}{{l^+l^-}}
\newcommand{\diel}{{e^+e^-}}
\newcommand{\dimu}{{\mu^+\mu^-}}
\newcommand{\dipi}{{\pi^+\pi^-}}
\newcommand{\dipiz}{{\pi^0\pi^0}}
\newcommand{\jpsi}{{J/\psi}}
\newcommand{\piz}{{\pi^0}}
\newcommand{\egcm}{{E_{\gamma}^*}}

\newcommand{\mee}{{M(\diel)}}
\newcommand{\mll}{{M(\dilep)}}

\newcommand{\gga}{{\gamma\gamma}}
\newcommand{\zze}{  {\dipiz\eta }}
\newcommand{\ppe}{  {\dipi\eta }}
\newcommand{\ppegg}{{\dipi\eta [\gga]}}
\newcommand{\tpi}{{\dipi\piz}}
\newcommand{\fpi}{{2(\dipi)}}
\newcommand{\spi}{{3(\dipi)}}
\newcommand{\vpi}{{2(\dipi)\piz}}
\newcommand{\fpz}{{\dipi 2\piz}}
\newcommand{\tpz}{{3\pi^0}}
\newcommand{\ppg}{{\pi^+\pi^-\gamma}}
\newcommand{\rg}{{\rho^0\gamma}}
\newcommand{\ree}{{\dipi\diel}}
\newcommand{\rmm}{{\dipi\dimu}}
\newcommand{\rll}{{\dipi\dilep}}
\newcommand{\meta}{M_{\eta}}
\newcommand{\metap}{M(\etap)}
\newcommand{\inv}{{\cal I}} %invisible}

\newcommand{\Mpp}{M_\psip}
\newcommand{\Mjp}{M_\jpsi}

\newcommand{\bx}{{{\cal B}(\etap\to X)}}
\newcommand{\bppe}{{{\cal B}(\etap\to\ppe)}}
\newcommand{\bppegg}{{{\cal B}(\etap\to\ppegg)}}
\newcommand{\begg}{{{\cal B}(\eta\to\gga)}}
\newcommand{\bb}{{\bppe\times\begg}}
\newcommand{\btpi}{{{\cal B}(\etap\to\tpi)}}
\newcommand{\btpz}{{{\cal B}(\etap\to\tpz)}}
\newcommand{\bzze}{{{\cal B}(\etap\to\zze)}}
\newcommand{\bree}{{{\cal B}(\etap\to\ree)}}
\newcommand{\brmm}{{{\cal B}(\etap\to\rmm)}}
\newcommand{\bdpi}{{{\cal B}(\etap\to\dipi)}}
\newcommand{\bfpi}{{{\cal B}(\etap\to\fpi)}}
\newcommand{\bvpi}{{{\cal B}(\etap\to\vpi)}}
\newcommand{\bspi}{{{\cal B}(\etap\to\spi)}}
\newcommand{\bfpz}{{{\cal B}(\etap\to\fpz)}}
\newcommand{\binv}{{{\cal B}(\etap\to\inv)}}

\newcommand{\einv}{{E_\gamma^*}}

\newcommand{\kev}{\,\mathrm{keV}}
\newcommand{\mev}{\,\mathrm{MeV}}
\newcommand{\gev}{\,\mathrm{GeV}}
\newcommand{\chivdof}{\chi^2_V/\mathrm{d.o.f.}}
\newcommand{\chimdof}{\chi^2_E/\mathrm{d.o.f.}}
\newcommand{\chimassdof}{\chi^2_M/\mathrm{d.o.f.}}
\newcommand{\gr}{\gamma_{\mathrm{r}}}
\newcommand{\ps}{P_{\mathrm{S}}}
\newcommand{\tg}{\gamma\gamma}
\newcommand{\gw}{\gamma\omega}
\newcommand{\pppz}{\dipi\piz}
\newcommand{\tptpz}{\dipi 2\piz}
\newcommand{\npi}{n\pi}
\newcommand{\eeg}{e^+e^-\gamma}
\newcommand{\eepp}{e^+e^-\dipi}
\newcommand{\eerho}{e^+e^-\rho^0}
\newcommand{\gtp}{\gamma\pppz}
\newcommand{\invis}{\ \mathrm{invisible}}

\preprint{CLNS~08/2040}       % for CLNS notes
\preprint{CLEO~08-22}         % for CLNS notes

\title{\Large \boldmath
Observation of $\etap$ decays to $\tpi$ and $\ree$}

\author{P.~Naik}
\author{J.~Rademacker}
\affiliation{University of Bristol, Bristol BS8 1TL, UK}
\author{D.~M.~Asner}
\author{K.~W.~Edwards}
\author{J.~Reed}
\author{A.~N.~Robichaud}
\author{G.~Tatishvili}
\affiliation{Carleton University, Ottawa, Ontario, Canada K1S 5B6}
\author{R.~A.~Briere}
\author{H.~Vogel}
\affiliation{Carnegie Mellon University, Pittsburgh, Pennsylvania 15213, USA}
\author{P.~U.~E.~Onyisi}
\author{J.~L.~Rosner}
\affiliation{Enrico Fermi Institute, University of
Chicago, Chicago, Illinois 60637, USA}
% ...
\collaboration{CLEO Collaboration}
\noaffiliation

\date{September 15, 2008}

\begin{abstract}

Using $\psip$$\to$$\dipi\jpsi$, $\jpsi\to\gamma\etap$ 
events acquired with the CLEO-c detector
at the CESR $\diel$ collider,
we make the first observations of the decays $\etap\to\tpi$
and $\etap\to\ree$, measuring absolute branching fractions 
$( 37^{+11}_{-\ 9}\pm 4)\times 10^{-4}$ and
$( 25^{+12}_{-\ 9}\pm5)\times 10^{-4}$, 
respectively.
For $\etap\to\tpi$, this result probes the 
mechanism of isospin violation and the roles of 
$\piz/\eta/\etap$-mixing and final state rescattering in strong decays.
We also set upper limits on branching fractions for $\etap$ decays 
to $\rmm$, $\fpi$, $\fpz$,  $\vpi$, $\spi$, and
invisible final states.

\end{abstract}


\maketitle

   Four decades after the first observation of the 
$\etap$ meson, its decays continue to provide a useful 
laboratory for probing strong interactions and new physics. 
Theoretical and experimental interest remains robust, 
in part because some expected modes have not
yet been observed at all and some rare or forbidden modes have
not been adequately limited. For example,
of all possible multi-pion $\etap$ decays, only $\etap\to 3\piz$ 
has been observed~\cite{alde}, and branching fraction limits for 
others are not stringent, lying in the range of 
(1-9)\%~\cite{notpub,danburg,ritten,ves}. 
No $\etap$ decays with an $\diel$ in the final state have been seen,
and just one with a dimuon ($\etap\to\gamma\dimu$) has been measured.
New physics would be indicated by invisible decays $\etap\to\inv$; 
\ie, decays that leave no trace in any detector because they
are composed of weakly interacting particles such as 
light dark matter. BES~\cite{besinv} has set the only such limit, 
$\binv<14\times 10^{-4}$ at 90\% confidence level (C.L.).

  Decay rates for three-pion decays of $\etap$ are commonly
expressed relative to their respective $\pi\pi\eta$ branching fractions
because they could arise from $\eta$-$\piz$ mixing:
$r_0\equiv\btpz/\bzze = (75\pm 13)\times 10^{-4}$~\cite{PDG2008} and
$r_\pm\equiv\btpi/\bppe$. 
The decay $\etap\to\tpi$ has garnered attention~\cite{ves,wasacosy,kloe,mamic}
both because experimental limits~\cite{notpub,danburg,ritten,ves} 
are large and because its rate can probe 
isospin symmetry breaking.
Under the two assumptions that $\tpi$ appears only through 
$\etap\to\dipi\eta$ followed by $\eta-\piz$ mixing
and that such decays populate the available phase space
uniformly, $r_\pm$ is found to be proportional 
to the light quark mass difference $(m_u-m_d)$~\cite{GTW}
and implies $r_\pm/r_0\simeq 0.37$~\cite{BMN}.
Suggesting neither assumption is justified, Ref.~\cite{BMN}
employs the framework of $U(3)$ chiral effective field
theory~\cite{BNLONG} to examine $\etap$ decays.
The incorporation of measured~\cite{vespipieta} 
$\etap\to\pi\pi\eta$ Dalitz slope parameters implies
a large contribution to $\etap\to\tpi$ from final state rescattering:
the prediction is that $r_\pm/r_0\simeq 5$ and that
dramatic structure should be present in the $\tpi$ Dalitz plot.

  Branching fractions for $\etap\to\dilep X$ 
($\ell^\pm\equiv e^\pm, \mu^\pm$) are expected to scale with
those for $\etap\to\gamma X$;
the most copious dileptonic decay should be $\etap\to\ree$.
Since other $\piz$ and $\eta$ decays to $\diel X$
occur at $\simeq$1\%~\cite{PDG2008} of the corresponding $\gamma X$ decay,
$\bree\simeq 0.3\%$ is expected. 
Two different theoretical approaches~\cite{FFK,BN}
both predict 
$\bree\simeq0.2$\%, $\rho^0$-dominance for the $\dipi$,
and an $\diel$ mass distribution peaking just above 2$m_e$
but with a long tail extending to $\simeq$300~MeV.
The experimental limit is $\bree<0.6\%$~\cite{RK}.
The corresponding dimuon channel is expected to be much rarer,
with predictions of $\brmm\simeq 2\times 10^{-5}$~\cite{FFK,BN}
and no experimental limit extant.

  In this Letter we search for decays of the $\etap$ meson
to eight final states: $\tpi$, $\ree$, $\rmm$, $\fpi$, $\spi$,
$\vpi$, $\fpz$, and $\inv$. 
Yields are normalized using the 
well-established decay chain $\etap\to\ppe$, $\eta\to\gga$, hereafter
denoted as $\etap\to\ppegg$, which was successfully used in a recent 
CLEO measurement~\cite{metap} of the $\etap$ mass
and found to provide a virtually background-free event sample. 
Events were acquired 
at the CESR $e^+e^-$ collider with the CLEO detector~\cite{CLEO}, 
mostly in the CLEO-c configuration (95\%) with the balance
from CLEO~III. The data sample corresponds to 27$\times$10$^6$~\cite{xnext} 
produced $\psip$ mesons, of which about  4$\times$10$^4$
decay as $\psip$$\to$$\dipi\jpsi$, $\jpsi\to\gamma\etap$. 

  For all the exclusive decay modes (\ie, all
but invisible), event selection requires 
finding every particle in the decay.
The tracking system must find exactly two 
oppositely-charged particles for the transition dipion,
and two, four, or six more tracks of net charge zero,
allowing for multiple combinations per event (which tends not to occur).
Photon candidates must have energy exceeding 37~MeV, and either be more
than 30~cm from any shower associated with one of the charged pions,
or, when between 15 and 30~cm from such a shower,
have a photon-like lateral shower profile. Showers are
rejected as photon candidates if they
lie near the projection of any charged pion's trajectory into
the calorimeter, or align with the initial momentum 
of any $\pi^\pm$ candidate within 100~mrad.
Photon candidates are ordered by energy, with the most
energetic always taken as the radiative photon from the $\jpsi$,
and subsequent ones, if required, must be taken as from the $\etap$.
That is, a shower can be included in the decay chain only
if every other photon of higher energy has also been used.
Photon pairs are candidates for a $\piz$ or $\eta$ if their
invariant mass satisfies $M(\gga)$=115-150~MeV or 500-580~MeV, respectively, 
and are then constrained to the known $\piz$ or $\eta$
masses~\cite{PDG2008}.

All decay products are constrained to originate 
from a single point (vertex) consistent with the beam spot.
The vertex-constrained event is additionally
constrained to the known $\psip$ mass~\cite{PDG2008} 
and three-momentum, including the
effect of the $\simeq$3~mrad crossing angle of the $e^+$ and $e^-$ beams.
Quality restrictions are applied to both the
vertex ($\chivdof<10$) and full event four-momentum ($\chimdof<10$) 
kinematic fits. 
From this point onward, all selections are based
upon the four-momenta obtained from the kinematic fit so
as to improve resolutions.
The mass recoiling against the $\psip$-to-$\jpsi$ transition
dipion must lie in the range 3092-3102~MeV.
The invariant mass of the $\etap$ candidate,
$\metap$, must lie in the window 952-964~MeV.
For the exclusive modes
[$\etap\to\inv$], sidebands in $\metap$ [$\einv$] are 
used to extrapolate a linear background level into the signal region.
Sideband intervals are, for $\metap$, 916-940 or 976-1000~MeV,
and, for $\einv$, 1220-1320 or 1460-1560~MeV. 

Candidates for $\ree$ are additionally required to have an $\diel$ invariant
mass $\mee$ below 100 MeV and which lies outside a window 
of 8-25~MeV, as well as to pass a tighter vertexing criterion, 
$\chivdof<3$. These
restrictions act to suppress
feed-across from $\etap\to\ppg$ when the photon converts
in the material in the vacuum pipe or detectors.
In such events the conversion electrons vertex poorly with the
other tracks and the beam spot. When forced to form a common vertex
with other tracks,
$\mee$ tends to be in the window 8-25~MeV 
due to the the discrete locations of the material.
Similar restrictions were used effectively
in Ref.~\cite{etabr} in the selection of $\eta\to\gamma\diel$ events.

No lepton identification is required for
$\ree$ or $\rmm$ candidate events. Instead,
all combinations of pion and lepton
mass assignments are made to the four charged
particles assigned to be the $\etap$ decay products,
and only those satisfying the respective kinematic fit are retained.
Background from $\ppg$ conversions with incorrectly swapped mass assignments
(\ie, when a pion is mistakenly assigned the electron mass and
 an electron the pion mass) 
are suppressed by the $\mee<100$~MeV requirement;
the $\mee =8-25$~MeV veto eliminates
conversion background with correct mass assignments.

Candidates for $\etap\to\tpi$ are additionally required to pass
a more restrictive criterion for the four-momentum fit
of $\chimdof<3$ to suppress feed-across from $\etap\to\ppg$,
which fakes $\tpi$ when a shower from 
a pion interaction in the calorimeter is erroneously taken as a 
photon candidate and happens to form a 
$\piz$ candidate with the real photon from the $\etap$ decay. 
For $\fpz$, the four-momentum fit must have
$\chimdof<3$ to suppress background from other $\jpsi$ decays.
For $\etap\to (2,3)(\dipi)$, the photon from
the $\jpsi$ decay may not pair with any other photon 
candidate in the event to form a $\piz$ or $\eta$ so as to suppress
backgrounds from $\jpsi\to (2,3)(\dipi)(\piz/\eta)$. 
To reduce feedacross from $\etap\to\ppe[\tpi]$ into $\etap\to\vpi$, 
candidates must not contain a three-pion combination
that satisfies a constraint to the $\eta$ mass 
with $\chimassdof<10$.
The sum of all unused photon
candidates' energies cannot exceed 75~MeV for $\fpz$ in order to suppress
backgrounds with higher neutral multiplicities.

  Candidates for $\inv$ decays are subject to a simpler set
of criteria. Exactly two charged particles of opposite charge
can be found in the event, and their recoil mass must
lie in the same window as the exclusive decays. 
Signal events would have a monochromatic photon
in the $\jpsi$ rest frame, so we require that the most
energetic photon candidate must, when boosted into the
$\jpsi$ center-of-mass using the dipion momentum, 
have energy $\einv$=1340-1440~MeV.
The sum of all unused photon candidates' energies must be 
less than 75~MeV.
The above restrictions on excess charged and neutral energy
are evaded by events in which the particles recoiling 
against the transition dipion and radiative photon do not
enter the active fiducial volume of the detector; hence
we require that
the missing momentum must have $|\cos\theta|<0.7$, assuring
the rejection of such events.
Background from $\jpsi\to\bar{n}n$ in which the 
neutron is undetected and the anti-neutron shower 
has energy in the signal window is suppressed by
requiring the radiative photon to have a lateral
profile consistent with that of an electromagnetic shower.

  Efficiencies for signal and feed-across from
other $\etap$ decays are modeled 
with Monte Carlo (MC) samples that were generated using the
{\sc EvtGen} event generator~\cite{evtgen},
fed through a {\sc Geant}-based~\cite{geant} detector simulation,
and subjected to event selection criteria. 
All $\etap$ decays are generated using phase space,
except that for $\rll$ and $\ppg$ we assume the dipion to come from a $\rho^0$,
and the $\mll$ distributions have been tuned 
to match those of Ref.~\cite{BN}. For invisible decays
we use $\etap\to\nu\bar{\nu}$.

\begin{figure}[t]
\includegraphics*[width=\figwid]{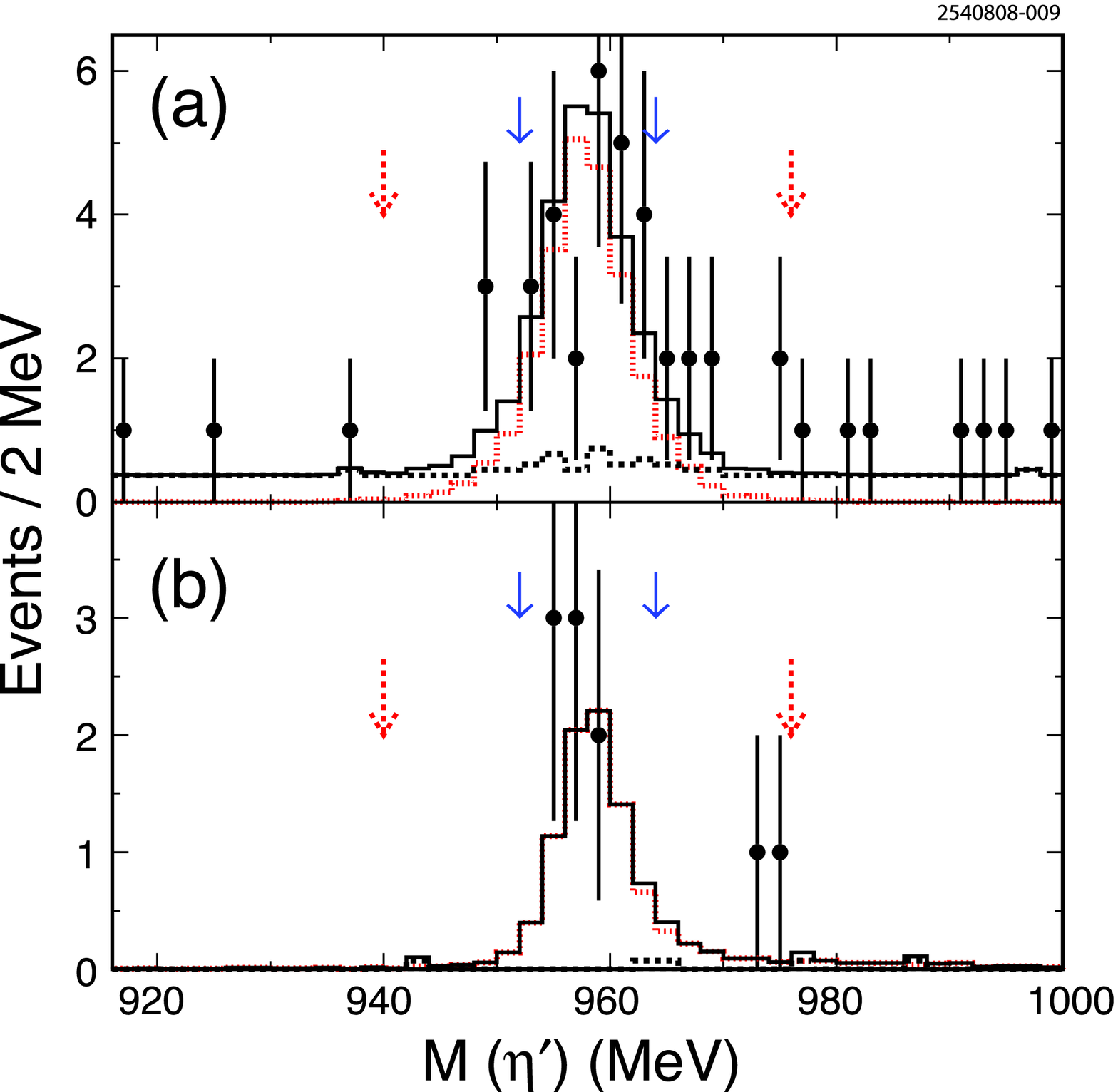}
\caption{Distributions in $\metap$ for (a) $\etap\to\tpi$
and (b) $\etap\to\ree$. Solid circles represent data
(nonzero bin entries only), 
the dashed histogram is the sum of a linear background normalized 
to the sideband populations in data and feed-across from $\etap\to\ppg$ 
normalized by branching fraction, 
the dotted histogram is the signal MC shape normalized to the observed signal level, 
and the solid line is the sum of dotted and dashed histograms.
Solid (dashed) arrows indicate nominal signal (sideband) region
boundaries;
sidebands extend to the edges of the plots.
All selection criteria are applied here except to $\metap$.
\label{fig:m3piree} }
\end{figure}

The data exhibit signals for $\etap\to\tpi$ (24 events) 
and $\etap\to\ree$ (8 events), 
with predicted background
levels of 3.8 and 0.14 events, respectively.
Distributions in $\metap$ appear in Fig.~\ref{fig:m3piree}. 
The signal level and corresponding 68\%~C.L.~interval 
in each case are obtained by 
subtracting the estimated background and accounting for
the statistics of signal- and sideband-region data as well as 
that of $\etap\to\ppg$ MC samples 
using a procedure similar to that of
Ref.~\cite{feldcous}. We consider two sources of
background, one peaking in the signal region 
(from other $\etap$ decays, in these two cases 
the only significant channel being $\etap\to\ppg$,
normalized by branching fractions~\cite{PDG2008} 
relative to $\etap\to\ppegg$)
and the second linear across the mass region. The
former is estimated from a MC sample to be 1.3 events
for $\tpi$ and 0.14 events for $\ree$,
and the latter from the mass sidebands to be 2.5 and 0 events,
respectively. For $\ree$, the two events between the
signal and sideband regions are consistent with tails of the 
signal. 

Statistical significance for each signal
is obtained from a large ensemble of simulated trials in which 
the backgrounds are thrown as appropriately-scaled Poisson 
distributions 
and the fraction of such trials in which the number of
events meets or exceeds that of the data is determined.
In both cases the significances exceed 6$\sigma$.

The number of events in the normalization mode, $\etap\to\ppegg$,
is evaluated in an identical manner as our signal modes,
and has no appreciable peaking feedacross background;
non-peaking backgrounds lead to a 0.2\% overall subtraction.
The absolute number of $\etap\to\ppegg$ events is
compatible with that expected from the size of
our data sample, the MC efficiency for this mode, and 
PDG branching fractions~\cite{PDG2008}.

\begin{figure}[t]
\includegraphics*[width=\figwid]{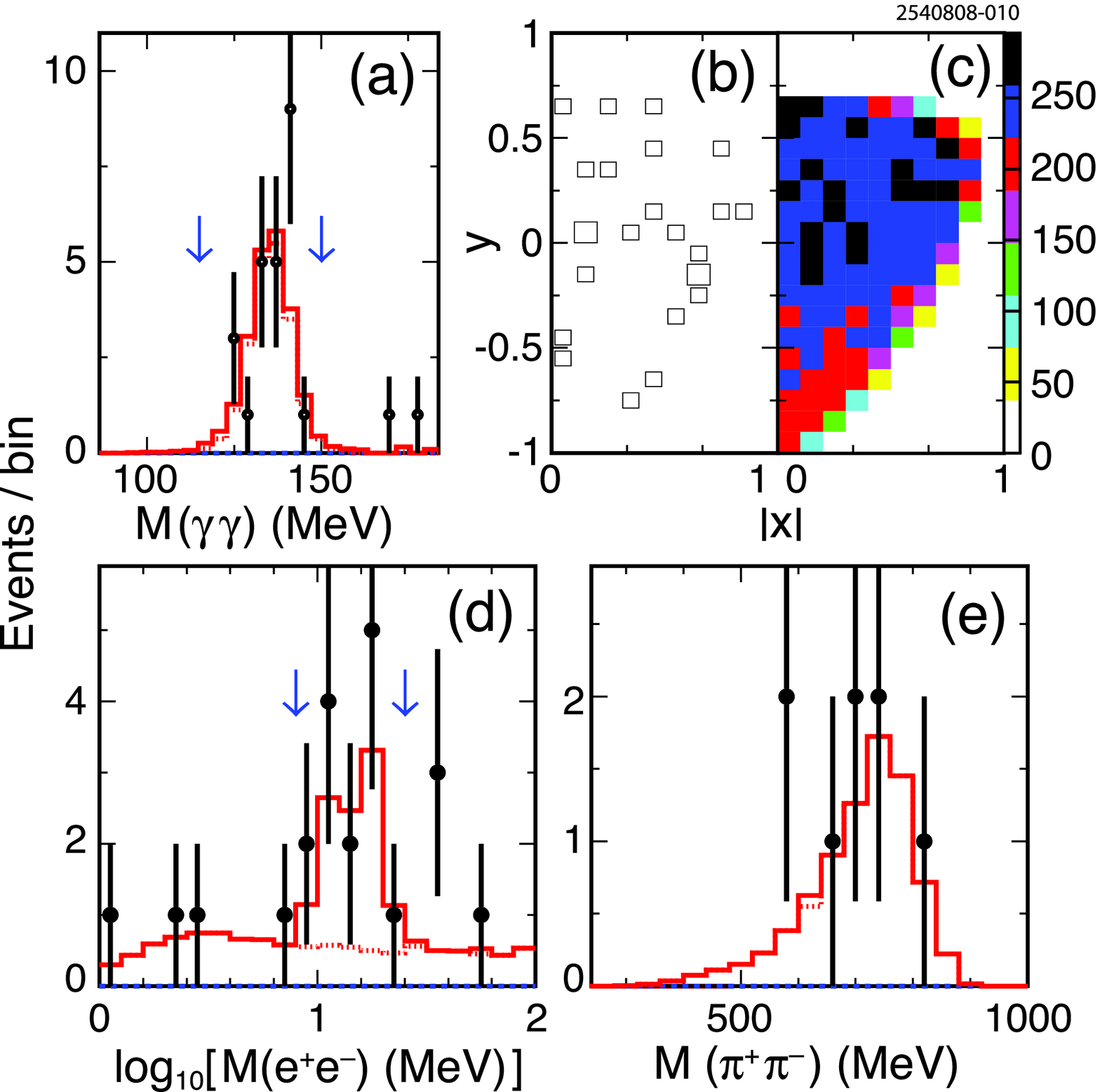}
\caption{Distributions in: (a) the invariant mass of the 
two photons in the $\piz$ candidate in $\etap\to\tpi$; 
(b) Dalitz variables $y$ vs.~$|x|$ for $\etap\to\tpi$,
uncorrected for efficiency,
for data, where box absence or size indicates 0, 1,
or 2 events in each 0.1-by-0.1 bin, and (c) from a phase-space
MC simulation, where bin shading indicates relative population;
(d) the $\diel$ invariant mass for $\etap\to\ree$;
(e) the $\dipi$ invariant mass 
for $\etap\to\ree$. In (a), (d), and (e),
solid circles represent the
data, the dotted histogram is the MC signal shape
normalized to the yields found in Table~\ref{tab:tableres}, 
and the solid line is the sum of MC signal 
and predicted $\etap\to\ppg$ feedacross.
The region between the arrows indicates the selected
region in (a) and an excluded region in (d).
All selection criteria, including that upon $\metap$, are applied here,
except to $M(\gga)$ in (a) and to $\mee$ in (d).
The quantities $x$ and $y$ are defined as $x\equiv \sqrt{3}(T_+-T_-)/Q$,
$y\equiv (3T_0/Q) -1$, $T_0$ $(T_\pm)$ is the $\piz$
($\pi^\pm$) kinetic energy in the $\etap$ center of mass, and 
$Q\equiv T_0+T_++T_-$. 
\label{fig:x3piree} }
\end{figure}

\begin{figure}[t]
\includegraphics*[width=\figwid]{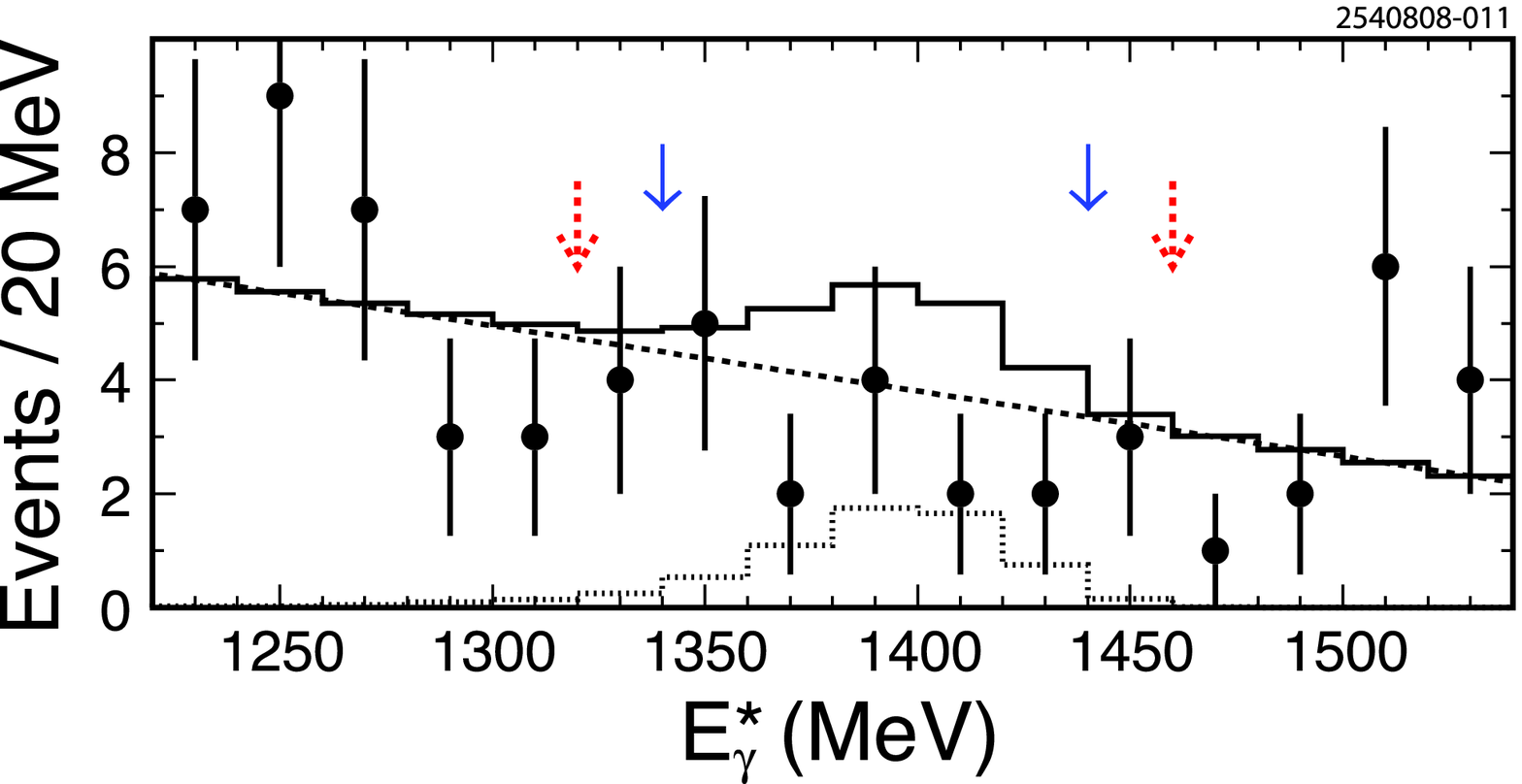}
\caption{Distribution of $E_\gamma^*$ (see text) 
for $\jpsi\to\gamma\etap$, $\etap\to\inv$ 
candidate events.
 Solid circles represent data
(nonzero bin entries only), 
the dashed histogram is the linear background normalized 
to the sideband populations in data, 
the dotted histogram is the signal MC shape normalized to the 
the 90\%~C.L. upper limit, 
and the solid line is the sum of dotted and dashed histograms.
Solid (dashed) arrows indicate nominal signal (sideband) region
boundaries;
sidebands extend to the edges of the plot.
All selections are applied here except to $E_\gamma^*$.
\label{fig:einv} }
\end{figure}

 Figure~\ref{fig:x3piree} shows kinematic
distributions for $\etap\to\tpi$ and
$\ree$;  within the statistical precision of so few 
events, we observe consistency of the data with the MC predictions.
In Fig.~2(a) the $\gamma\gamma$ mass distribution
for the $\piz$ candidate in $\etap\to\tpi$ verifies the 
cleanliness of that sample.
Of particular interest for the $\etap\to\tpi$ decay is the 
Dalitz plot distribution shown in 
Fig.~\ref{fig:x3piree}(b) and (c), 
where data and phase-space MC simulation are shown side by side
(these can be compared to the prediction in Fig.~1 of Ref.~\cite{BMN}).
We compare the Dalitz plot population density of data points to
the two predictions and find much better agreement with 
the phase-space model than with that of rescattering 
through $\rho^\pm$~\cite{BMN}.
The $\einv$ distribution for $\etap\to\inv$ appears in 
Fig.~\ref{fig:einv} and shows no indication of a signal.
For the channels where no signals are apparent, we compute
90\%~C.L.~upper limits on the signal yields.

Table I displays the numerical results for each mode.
No data events within the $\metap$ signal window are seen
for the modes $\dipi\dimu$, $2(\dipi)$, or $3(\dipi)$;
for $2(\dipi)\piz$ we find one signal event with 0.5 background,
and for $\dipi 2\piz$, 5 signal events with 2 background.
The production rate $R$ of each mode relative to that of 
$\etap\to\ppegg$ is defined as 
$R\equiv[ \bx/\bb ]$,
and the absolute branching fraction $B\equiv\bx$.
$R$ is obtained by dividing the yield by its
efficiency relative to the normalization mode
and the number of events in the normalization mode;
$B$ is obtained from $R$ by multiplying it by  ($0.1753\pm  0.0056$),
the value of the denominator in 
$R$
using
branching fractions compiled in Ref.~\cite{PDG2008}.
Overall normalizations cancel in the values of $R$,
as do some of the track- and photon-finding systematic
errors, depending upon mode. Systematic errors
include detector modeling, the background linearity assumption,
and the possible presence of intermediate resonances,
amounting to 10-20\%, depending upon mode; 
however, statistical errors dominate
the systematic uncertainties here. The final
column shows previous measurements, if any, for each mode:
our measurements provide the first limits for 
$\etap\to\rmm$ and $\fpz$ and improve upon
those for the other modes.

In conclusion, we report the first observation of
the decays $\etap\to\tpi$ and $\etap\to\ree$ and 
measurement of their branching fractions.
We find $\btpi = ( 37^{+11}_{-\ 9}\pm 4)\times 10^{-4}$ and
 $r_\pm=(83\pm 22)\times 10^{-4}$.
Using the branching fractions of Ref.~\cite{PDG2008}, 
we determine $r_\pm/r_0=1.11\pm 0.35$, 
more than 2 standard deviations above
the $\piz$-$\eta$-mixing prediction of 0.37, 
and far below the chiral unitary framework 
prediction of 5~\cite{BMN}.
The dileptonic results $\bree = ( 25^{+12}_{-\ 9}\pm5)\times 10^{-4}$ 
and $\brmm < 2.4\times 10^{-4}$ are consistent
with predictions~\cite{FFK,BN}.
We also obtain first or improved branching fraction upper limits for 
$\etap$ decays to multi-pion and invisible final states.

We gratefully acknowledge the effort of the CESR staff
in providing us with excellent luminosity and running conditions.
This work was supported by
the A.P.~Sloan Foundation,
the National Science Foundation,
the U.S. Department of Energy,
the Natural Sciences and Engineering Research Council of Canada, and
the U.K. Science and Technology Facilities Council.

\clearpage

\begin{table}[t]
\setlength{\tabcolsep}{0.30pc}
\catcode`?=\active \def?{\kern\digitwidth}
\caption{Results for $\etap\to X$ search, showing for
each mode $X$ the efficiency relative to that of
the normalization mode $\etap\to\ppegg$,  $\epsilon/\epsilon_0$;
the net number of signal events,
after background subtractions, $N$ (or 90\%~C.L.~upper limit where indicated
with ``$<$''); the branching fraction ratio $R$ [see text]; 
the absolute branching fraction $B\equiv\bx$ and its previous upper
limit $P$~\cite{PDG2008}. Entries
for $R$ and $B$ include systematic errors.
}
\label{tab:tableres}
\begin{center}
\begin{tabular}{lrcccc}
\hline
\hline
Mode $X$ & $\epsilon/\epsilon_0$\ & $N$ & $R (10^{-3})$ & $B (10^{-4})$ & $P (10^{-4})$ \\
\hline
$\ppegg$ & 1.00 & 1756$\pm$42          & - & - & -\\
$\tpi$   & 0.55 & 20.2$^{+6.1}_{-4.8}$ & 21$^{+6}_{-5}\pm$2   & 37$^{+11}_{-\ 9}\pm$4
& $<$500 \\
$\ree$   & 0.31 &  7.9$^{+3.9}_{-2.7}$ & 14$^{+7}_{-5}$$\pm$3 & 25$^{+12}_{-\ 9}$$\pm$5
& $<$60 \\
$\rmm$   & 2.14 & $<$4.8               & $<$1.3 & $<$2.4 & - \\
$\fpi$   & 1.02 & $<$2.3               & $<$1.4 & $<$2.4 & $<$100 \\
$\fpz$   & 0.18 & $<$4.1               & $<$15  & $<$27  & - \\
$\vpi$   & 0.21 & $<$3.6               & $<$11  & $<$20  & $<$100 \\  
$\spi$   & 0.47 & $<$2.3               & $<$3.0 & $<$5.3 & $<$100 \\
Invisible& 0.74 & $<$5.8               & $<$5.4 & $<$9.5 & $<$14 \\  
\hline
\hline
\end{tabular}
\end{center}
\end{table}

\clearpage

\begin{thebibliography}{99}

\bibitem{alde} D.~Alde \etal, Yad. Fiz. {\bf 47}, 385 (1988)
               [Sov. J. Nucl. Phys. {\bf 47}, 243 (1988)]; 
               F.G.~Binon \etal, Phys. Lett. B {\bf 140}, 264 (1984).

\bibitem{notpub} Ref.~\cite{danburg} quotes $\btpi<9$\%.
                 Ref.~\cite{ritten} quotes $\btpi<5$\%, but is unpublished.
                 VES~\cite{ves} has reported an upper limit 
                 $\btpi<1.75$\%, but it is a preliminary conference result.

\bibitem{danburg} J.S.~Danburg \etal, Phys. Rev. D {\bf 8}, 3744 (1973).
 
\bibitem{ritten} A.~Rittenberg, Ph.D. Thesis 1969 (unpublished),
\\ 
%\begin{verbatim}
{\tt http://repositories.cdlib.org/lbnl/UCRL-18863  }
%\end{verbatim}

\bibitem{ves} V.~Nikolaenko \etal\ (VES Collaboration), AIP Conf. Proc. 
              {\bf 796}, 154 (2005).

\bibitem{besinv} M.~Ablikim \etal\ (BES Collaboration), Phys. Rev. Lett. {\bf 97}, 202002 (2006).

\bibitem{PDG2008} C. Amsler \etal\ (Particle Data Group), Phys. Lett. B
  {\bf 667}, 1 (2008).

\bibitem{wasacosy} H.H.~Adam \etal\ (WASA-at-COSY Collaboration),
                   arXiv:nucl-ex/0411038 (2004);
                  M.J.~Zielinski, arXiv:0807.0576v1 [hep-ex] (2008).

\bibitem{kloe} C.~Bloise, AIP Conf. Proc. {\bf 950}, 192 (2007).

\bibitem{mamic} A.~Thomas, AIP Conf. Proc. {\bf 950}, 198 (2007).

\bibitem{GTW} D.J.~Gross, S.B.~Treiman, and F.~Wilczek, Phys. Rev. D
  {\bf 19}, 2188 (1979).

\bibitem{BMN} B.~Borasoy, U.-G.~Meissner, and R.~Nissler,
  Phys. Lett. B {\bf 643}, 41 (2006).

\bibitem{BNLONG} B.~Borasoy and R.~Nissler, Eur. Phys. J. A {\bf 26}, 383 (2005).

\bibitem{vespipieta} V. Dorofeev \etal\ (VES Collaboration), Phys. Lett. B {\bf 651}, 22 (2007). 

\bibitem{FFK} A.~Faessler, C.~Fuchs, and M.I.~Krivoruchenko,
  Phys. Rev. C {\bf 61}, 035206 (2000).

\bibitem{BN} B.~Borasoy and R.~Nissler, Eur. Phys. J. A {\bf 33}, 95 (2007).

\bibitem{RK} A.~Rittenberg and G.R.~Kalbfleisch, Phys. Rev. Lett. {\bf 15}, 556 (1965).

\bibitem{metap} J.~Libby \etal\ (CLEO Collaboration), Phys. Rev. Lett
  {\bf 101}, 182002 (2008).

\bibitem{CLEO} 
  Y.~Kubota \etal\  (CLEO Collaboration),
  Nucl.\ Instrum.\ Meth.\ A {\bf 320}, 66 (1992);
  M.~Artuso \etal,
  Nucl.\ Instrum.\ Meth.\ A {\bf 554}, 147 (2005);
  D.~Peterson \etal,
  Nucl.\ Instrum.\ Meth.\ A {\bf 478}, 142 (2002);
  CLEO-c/CESR-c Taskforces \& CLEO-c Collaboration, 
  Cornell University LEPP Report No.
  CLNS 01/1742, 2001 (unpublished).

\bibitem{xnext} H.~Mendez \etal\ (CLEO Collaboration), Phys. Rev. D {\bf 78}, 011102 (2008).

\bibitem{etabr} A.~Lopez \etal\ (CLEO Collaboration), Phys. Rev. Lett. {\bf 99}, 122001 (2007).

\bibitem{evtgen} D.J. Lange, Nucl. Instrum. Methods Phys. Res., Sect. A {\bf 462}, 152 (2001). 

\bibitem{geant} R. Brun \etal, 
  {\textsc Geant} 3.21, CERN Program Library Long Writeup W5013, 1993 (unpublished). 

\bibitem{feldcous} G.J.~Feldman and R.D.~Cousins, Phys. Rev.D {\bf 57},  3873 (1998). 


\end{thebibliography}
\end{document}